\documentclass{article}
\usepackage{graphicx} % Required for inserting images

\usepackage{arxiv}

\usepackage{amssymb}
\usepackage{amsthm}
\usepackage{amsmath}
\usepackage{dsfont}
\usepackage{xcolor}
\usepackage{booktabs} % nice tabular
\usepackage{graphicx}
\usepackage{appendix}
\usepackage{bm}
\usepackage{algorithm}
\usepackage{algpseudocode}
\usepackage{adjustbox}
\usepackage{caption}
\usepackage{subcaption}
\usepackage{soul}

\usepackage{everypage}

\newcommand{\R}{\mathds{R}}
\newcommand{\C}{\mathds{C}}

\newcommand{\alphab}{{\vec{\alpha}}}

\newcommand{\tcb}[1]{#1}

\usepackage{todonotes}

\title{A practical approach to determine minimal quantum gate durations using amplitude-bounded quantum controls}

\date{November 8, 2023}
\author{ Stefanie G{\"u}nther\thanks{Corresponding author}, ~ N.~Anders Petersson \\
	Center for Applied Scientific Computing\\
 Lawrence Livermore National Laboratory, Livermore CA 94550\\
	\texttt{\{guenther5, petersson1\}@llnl.gov} 
}

\renewcommand{\headeright}{}
\renewcommand{\undertitle}{}

\begin{document}

\maketitle

\begin{abstract}
We present an iterative scheme to estimate the minimal duration in which a quantum gate can be realized while satisfying hardware constraints on the control pulse amplitudes. 
The scheme \tcb{performs a sequence of unconstrained numerical optimal control cycles that each minimize the gate fidelity for a given gate duration alongside an additional penalty term for the control pulse amplitudes. 
After each cycle, the gate duration is adjusted based on the inverse of the resulting maximum control pulse amplitudes, by re-scaling the dynamics to a new duration where control pulses satisfy the amplitude constraints. Those scaled controls then serve as an initial guess for the next unconstrained optimal control cycle, using the adjusted gate duration.
}
We provide multiple numerical examples that each demonstrate fast convergence \tcb{of the scheme} towards a gate duration that is close to the quantum speed limit, given the control pulse amplitude bound.
\tcb{The proposed technique is agnostic to the underlying system and control Hamiltonian models, as well as the target unitary gate operation, making the time-scaling iteration an easy to implement and practically useful scheme for reducing the durations of quantum gate operations.}
\end{abstract}


\section{Introduction}
Quantum optimal control utilizes numerical optimization tools to design pulses that drive quantum devices. It's potential to improve fundamental operations in Quantum Information Science has by now been demonstrated on various applications, e.g., for quantum state preparation \cite{Rojan2014arbitrary},  applications to quantum error correction \cite{waldherr2014quantum, gaitan2008quantum}, and the realization of logical quantum gate operations \cite{doi:10.1080/09500340802344933, PhysRevLett.89.188301, cho2023direct}, see for example \cite{glaser2015training, koch2016controlling} for an overview.
In the current Noisy Intermediate-Scale Quantum (NISQ) era, it is desirable to design control pulses with minimal gate duration, such that the quantum operation can be performed before decoherent processes collapse the quantum information into a classical state.
However, finding the minimal duration in which such an operation is realizable is challenging. When control pulse amplitudes are unlimited, various lower bounds known as intrinsic \textit{quantum speed limits} (QSL) have been established for the time-scale over which a quantum system can evolve, see, e.g., the overview in \cite{deffner2017quantum}.
Such speed limits can be computed analytically for simple model problems \cite{hegerfeldt2013driving, khaneja2001timeoptimal, khaneja2002subriemannian}, and can be estimated numerically for more complex systems by performing multiple optimal control cycles, sweeping over a range of gate durations.
Indeed it has been demonstrated on specific examples that such limits align with the shortest durations for which numerical optimal control techniques converge, when control pulse amplitudes are unlimited \cite{hegerfeldt2013driving, caneva2009quantumspeedlimit, ashhab2012speed}. 
In practice, however, control pulses are generally subjected to hardware constraints, most commonly given in terms of a maximum drive strength as defined by hardware waveform generating devices. 
When hardware constraints bound the control pulse amplitude, the time-scale required to realize a quantum operation can be significantly slower than the theoretical (intrinsic) QSL. 
For unitary operations in the closed system setting, 
an extrinsic speed limit that takes the maximum control pulse amplitude into account has been derived in \cite{arenz2017roles}. 
While the lower bound matches the minimum gate duration remarkably well for small single-qubit cases, and the extension in \cite{lee2018dependence} to general N-level system demonstrates qualitatively reasonable scaling properties for larger numbers of qubits, the lower bound is not sharp in the sense that the actual gate duration is no longer precisely estimated as system sizes increase. 
Currently, a general technique to find minimal gate durations when control pulse amplitudes are bounded by hardware constraints is not available. 
Instead, it is common practice to determine the minimum gate duration by trial and error through multiple optimal control cycles, each using a different pulse duration and a different initial guess for the control pulse. Henceforth, the latter approach will be referred to as the brute-force method.

In~\cite{seifert2022time}, a re-seeding technique was proposed that adjusts gate durations based on previous success or failure of the optimal control iterations, combined with a bisection of the resulting gate durations. 
However, imposing parameter bounds during the optimization updates often leads to non-convex optimization landscapes \cite{ge2022optimization} and slow or stagnant optimization progress, such that optimization cycles need to be performed multiple times for various different initial guesses.

In this paper, we propose an alternative scheme to numerically find minimal time-scales in which a unitary operation can be realized under amplitude-constrained control fields, by re-scaling the gate duration based on the optimized control pulse amplitude. Instead of directly imposing box constraints on the control pulse, we include a penalty term in the objective function that minimizes the control pulse energy during the optimization. After solving the unconstrained and penalized optimization problem for a given gate duration, we re-scale the dynamics to a new duration where control pulses satisfy the amplitude constraints. Those stretched (or squeezed) control pulses then serve as an initial guess for the next optimization cycle for the new gate duration.
We present numerical evidence for system of up to three coupled qudits where this iterative scheme rapidly converges to a final gate duration close to the QSL, yielding control pulses that satisfy hardware-specific amplitude bounds while minimizing the gate time duration. 

While the approach presented here is agnostic to the underlying Hamiltonian model and optimization strategy, in Section \ref{sec:optimalcontrol} we present the system and control Hamiltonian models that are used in our numerical experiments. Section \ref{sec:mintime} describes the proposed iterative scheme for finding minimal gate durations and corresponding amplitude-constrained control pulses. Section \ref{sec:results} presents the numerical results, and conclusions are drawn in Section \ref{sec:conclusion}.

\section{The quantum optimal control problem}\label{sec:optimalcontrol}
Given a unitary target gate $V^{\textrm{target}}$ that represents a logical quantum operation, the goal of quantum optimal control is to design optimal pulses that drive any initial quantum state $\phi(0)$ at time $t=0$ to the unitarily transformed target state $\phi(T)=V^{\textrm{target}}\phi(0)$ at time $T>0$. To that end, gradient-based optimization methods are typically applied to minimize a cost function that measures the distance between the target and the driven evolution in terms of the gate infidelity,
\begin{align}\label{eq:finaltime_cost}
    \min J_{cost}\left(U(T)\right) = 1-\left| \frac{1}{N} \mbox{Tr} \left(U^\dagger(T) V^{\textrm{target}}\right) \right|^2,
\end{align}
where the columns of $U(T)$ describe the evolved quantum state at final time $T$ for a basis of initial states $U(0) = [e_0,\dots, e_{N-1}]$, where $e_i$ are the canonical basis vectors. We here consider closed quantum systems modeled by Schroedinger's equation, such that the evolved dynamics satisfy
\begin{align}\label{eq:mastereq}
 \frac{d U(t)}{dt} = &-iH(t)U(t), \quad 0\leq t \leq T,\quad U(0) = I.
\end{align}
The Hamiltonian,
\begin{align}\label{eq:schroedinger}
    H(t) = H_{sys} + H_c(t), %\sum_{q=1}^Q f_q(t)H_q
\end{align}
decomposes into a time-independent system part, $H_{sys}$, and a time-varying control part, $H_c(t)$. In the laboratory-frame of reference, the control Hamiltonian is given by $\sum_q f_q(t)(a_q + a_q^\dagger)$, where the real-valued control pulse satisfies $f_q(t) = 2\, \mbox{Re}\{c_q(t) e^{i t \omega^{d}} \}$, where $\omega^d$ is the drive frequency. To slow down the time scales of the state dynamics and the control pulse, we apply the rotating wave approximation (RWA) and select the frequency of rotation to equal the drive frequency, $\omega^{rot} = \omega^d$.
In the rotating frame, we consider a general system Hamiltonian \tcb{for} modeling $Q\geq 1$ coupled superconducting qudits,
\begin{align}
  H_{sys} = \sum_{q=1}^Q &{\left(\omega_q - \omega^{rot}\right)} a_q^\dagger a_q - \frac{\xi_q}{2} a_q^{\dagger}a_q^{\dagger}a_q a_q \nonumber \\
  &+ \sum_{p>q} J_{pq} \left(a_p^\dagger a_q + a_pa_q^\dagger\right),
\end{align}
where $\omega_q$ denotes the 0-1 transition frequency of qudit $q$, $\xi_q$ is the self-Kerr coefficient, and $J_{pq}$ denotes the \tcb{dipole-dipole} coupling coefficient between qudits $p$ and $q$. Further, $a_q$ ($a_q^\dagger$) denote the lowering (raising) operator for qudit $q$. 
The action of external control pulses on qudit $q$ in the rotating frame is given by the control Hamiltonian
\begin{align}
   H_c(t) = \sum_{q=1}^Q c_q(t) a_q + c_q^*(t) a_q^\dagger.
\end{align} 
where $c_q(t) = p_q(t) + iq_q(t)$ denote the control pulse for qubit $q$ in the rotational frame and $c_q^*(t)$ denotes its complex conjugate. 
For notational simplicity, we now drop the index $q$ denoting each subsystem (qudit), noting that it is \tcb{straightforward} to apply the proposed method to systems with multiple qudits.

To parameterize the control pulse $c(t)$ in the rotating frame, we here choose B-spline basis functions:
\begin{align}\label{eq:bsplinebasis}
  c(t) &= \sum_{s=1}^{N_s}\alpha_{s} B_s(t),
\end{align}
where $\alpha_s = \alpha_s^{real} + i\alpha_s^{imag} \in \C$ are the control parameters that are being optimized for; $B_s(t)$ are fixed, quadratic, \tcb{cardinal} B-spline basis functions~\tcb{\cite{PiegTill96}} with local support in time.
In the following we will refer to the set of control parameters $\alphab := \{\alpha_s^{real}, \alpha_s^{imag}\}_{s=1}^{N_s}$ as the control vector. It determines the control pulse $c = c(t;\alphab)$ in the control Hamiltonian and hence the solution operator $U = U(t;\alphab)$ of Schroedinger's equation \eqref{eq:mastereq}.\footnote{In the following, the dependence on $\alphab$ will be suppressed for improved readability.}
Parameterizing the control functions using quadratic B-splines provides a compact alternative to discretizing the control functions on the same time step as the underlying dynamical system. \tcb{The cardinal quadratic B-spline basis functions result in control pulses that are continuously differentiable, see Appendix~\ref{app:bspline}}. 
\tcb{Therefore, they} can readily be applied to the control hardware, without the need for smoothing or filtering a stair step control pulse or a bang-bang ansatz.
\tcb{While B-splines have many attractive properties~\cite{Unser97}, we note that the time-scaling algorithm presented in the following section is not restricted to any particular class of pulse parameterization. The only requirement is that the parameterization must allow the pulses to readily be stretched or compressed to change the pulse duration.}

%%%%%%%%%%%%%%%%%%%%%%%%%%%%%%%%%%%%%%%%%%%%%%%%%%%%%%%%%
\section{Time-scaling iteration to find minimal gate duration under pulse amplitude constraints}\label{sec:mintime}

When optimal controls are applied in practice, the pulses are typically subject to amplitude bounds as specified by the pulse-generating hardware, such as an Arbitrary Waveform Generator (AWG). Typically those constraints are given in terms of a maximum drive strength $b_{max}$ such that optimized pulses need to satisfy
\begin{align}
    \max_{t\in[0,T]} |c(t)| \leq b_{max}.
\end{align}
While the amplitude bound varies depending on the specific AWG, we here consider a maximum amplitude bound of $b_{max}{/2\pi} = 40$ MHz for the numerical results presented below.\footnote{Because we have scaled Schroedinger's equation to make $\hbar = 1$, the unit of energy becomes angular frequency. As a result, the unit of the control pulse amplitude $|c(t)|$ is also angular frequency.}
Standard optimal control approaches incorporate such constraints into the optimization procedure by imposing box-constraints on the control parameter vector, e.g., by projecting the gradient-based update steps onto the given bounds, or by using interior point or barrier methods that augment the objective function and enforce the constraint though penalty terms \cite{nocedal2006numerical}. However, when control parameters hit the bounds, optimization convergence can deteriorate due to non-convex optimization landscapes \cite{russell2016quantum}. As a result, the constrained optimization problem is often much harder to solve than the unconstrained one. In practice, it is often required to repeat multiple optimization cycles, each for a different randomized initial guesses of the control parameters and different pulse durations, to increase the chances of finding a global optima.

In this section, we propose an iterative scheme to automatically adjust the gate duration based on the energy norm of the optimized control pulse.
Instead of explicitly imposing box-constraints on the control parameter vector during the optimization procedure, we add a penalty term to the objective function that aims to minimize the control pulse energy. The iterative scheme performs a sequence of such unconstrained optimal control iterations, each using an updated gate duration that is selected based on the control pulse energy in the previous iteration. 

In each iteration $k$ of the proposed scheme, we solve the following augmented and unconstrained optimal control problem for a given gate duration $T_k$, 
\begin{align}\label{eq:optimproblem}
    \min_{\alphab} & \quad J_{\textrm{infid}}\left(U(T_k;\alphab)\right) + \gamma J_{\textrm{energy}}(c(\tcb{\cdot};\alphab)).
\end{align}
Here, $\gamma>0$ is a parameter that weights the contribution of the control energy term, \begin{align}\label{eq:energynorm}
    J_{\textrm{energy}}(c(\tcb{\cdot}{;\alphab})) = \frac{1}{T_k}\int_0^{T_k} |c(t;\alphab)|^2 \, dt,
\end{align}
relative to the infidelity, $J_{\textrm{infid}}$, in the objective function.

\begin{figure}[htb]
    \centering
    \includegraphics[width=0.6\textwidth]{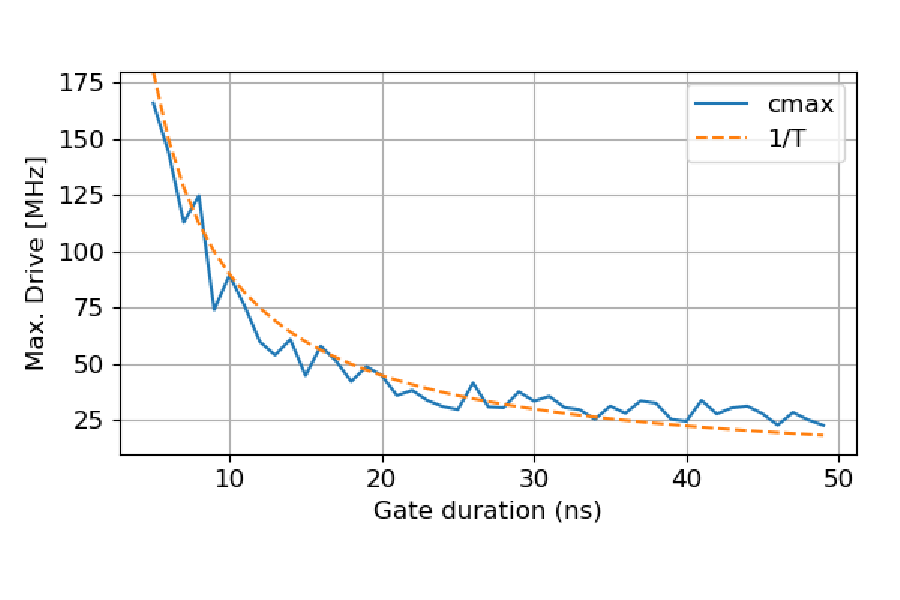}
    \caption{\tcb{Unconstrained minimization of the infidelity and the control energy \eqref{eq:optimproblem}, illustrating the relation between the maximum control amplitude and the gate duration $T$, compared with an $1/T$ scaling. Here, the target unitary is a $QFT$ gate on a 4-level qudit.}}
    \label{fig:H4_nobounds_dmax}
\end{figure}

Performing an optimization cycle that minimizes the weighted sum of the infidelity at the final time \eqref{eq:finaltime_cost} and the control pulse energy \eqref{eq:energynorm} yields control pulses with the smallest energy norm possible, while realizing the given target unitary at time $T_k$. 
We can assume that a gradient-based optimal control technique can solve this unconstrained optimization problem, given that the weight $\gamma$ is chosen appropriately. 
However, since box-constraints are not imposed during the optimization, the resulting control pulses, while being minimal in terms of their energy, might  exceed the given hardware bounds $b_{max}$.
Figure \ref{fig:H4_nobounds_dmax} demonstrates the relation between the pulse duration $T$ and the resulting maximum control amplitude
\begin{align}
    c_{max} = \max_{t\in [0,T]} |c_*(t)|,
\end{align}
for optimized unconstrained control pulses $c_*(t)$, on a 4-level qudit QFT gate (problem specifications are given in Section \ref{sec:results}).
\tcb{Here, each optimization proceeded until the gate infidelity dropped below the threshold $10^{-4}$. Note that the maximum control amplitude approximately scales as $1/T$.}

\begin{figure}[htb]
    \centering
    \includegraphics[width=0.5\textwidth]{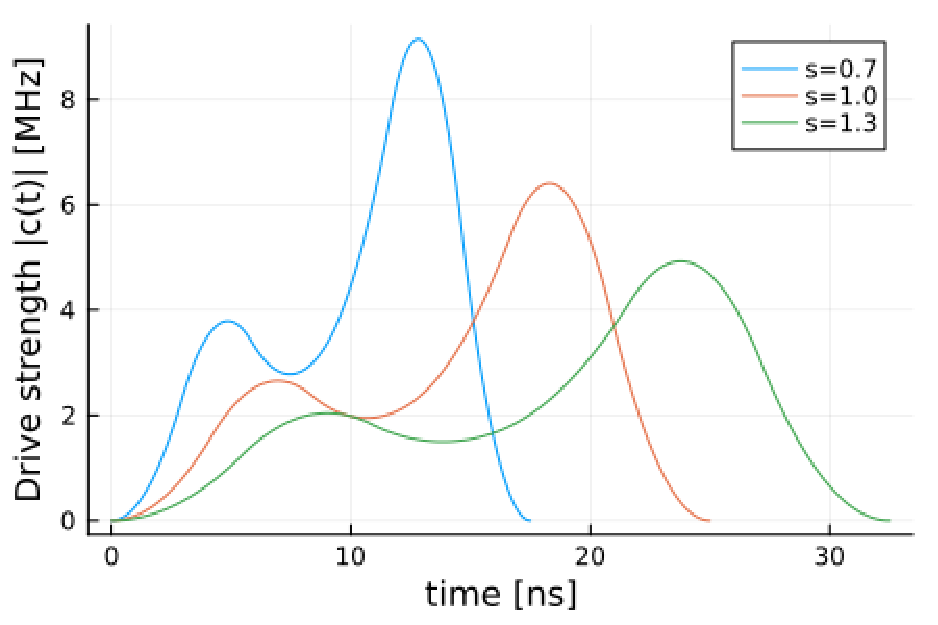}
    \caption{Re-scaling control pulses: By choosing the scaling parameter $s$, the maximum drive strength of the stretched ($s>1$) or compressed ($s<1$) control pulse $\tilde c(t) = \frac{1}{s}c(t/s)$ can be adjusted.}
    \label{fig:rescale}
\end{figure}

We make use of the scaling $c_{max}\sim 1/T$ to find a minimal gate duration for given amplitude constraints by scaling the current gate duration $T_k$ by the ratio between the current maximum control amplitude resulting from solving the unconstrained optimization problem \eqref{eq:optimproblem}, and the given hardware constraint $b_{max}$, i.e., we propose the update scheme 
\begin{align}\label{eq:scaling}
    \tcb{T_{k+1}} = sT_k, \quad \text{where}\quad s = \frac{c_{max}}{b_{max}}.
\end{align}
Consider the scaled time variable 
\begin{align}
\tau(t) = {st} \quad \Rightarrow \quad \tau \in [0,\tcb{T_{k+1}}] \quad \text{for} \quad t\in [0,T_k],
\end{align}
and corresponding scaled evolution $\tilde U(\tau) := U(\tau/s) = U(t)$. The scaled unitary $\tilde U(\tau)$ satisfies the scaled Schroedinger equation
\begin{align}
    \dot{\tilde U}(\tau) = \frac{d U(\tau/s)}{d \tau} = \frac{1}{s}\dot U(\tau/s) = -\frac{i}{s}H(\tau/s)\tilde U(\tau) \\
    \text{for} \quad \tau\in (0,\tcb{T_{k+1}}), \nonumber
\end{align}
and, hence, $\tilde U(\tau)$ evolves under the scaled Hamiltonian
\begin{align}
    \tilde H(\tau) := \frac{1}{s}H\left(\tau/s\right) = \frac{1}{s}H_{sys} + \tilde c(\tau)  a + \tilde c(\tau)^* a^\dagger
\end{align}
with the control pulse
\begin{align}
    \tilde c(\tau) := \frac{1}{s} c(\tau/s) 
\end{align} 
The scaled pulse stretches (if $s>1$) or compresses (if $s<1$) the original pulse into the time domain \tcb{$[0,T_{k+1}]$} while preserving its integral,
\begin{equation}
    \int_{0}^{T_k} |c(t)|\, \mathrm{d}t = \int_0^{T_{k+1}} |\tilde{c}(\tau) |\, \mathrm{d}\tau. 
\end{equation}
Preserving the control pulse integral can be further motivated by solving the control problem analytically for a model problem, see Appendix~\ref{app:displacement}. By parameterizing the control functions in terms of  B-spline basis functions as in \eqref{eq:bsplinebasis}, the re-scaled controls can be easily computed by scaling the parameter vector $\tilde {\alpha} = s^{-1}{\alpha}$, and re-evaluating the B-spline basis functions at \tcb{the scaled time}, %\st{$\tau/s$}
hence stretching or compressing their envelops, see Figure \ref{fig:rescale}.
By design (choice of the scaling parameter $s$ in \eqref{eq:scaling}), the scaled control pulses satisfy the hardware bounds exactly with 
\begin{align}
    \max_{\tau \in [0,\tcb{T_{k+1}}]} |\tilde c(\tau)| = b_{max}
\end{align}
Further, if the dynamics had obeyed the scaled system Hamiltonian $\frac{1}{s}H_{sys}$, the scaled controls would have realized the target gate at time \tcb{$T_{k+1}$}.
However, because the actual system Hamiltonian is $H_{sys}$, we only use the scaled control pulse as an initial guess to solve the optimization problem \eqref{eq:optimproblem} for the updated gate duration, \tcb{$T_{k+1}$}. This procedure is repeated until the scaling factor remains constant at $s=1$, indicating that the minimal gate duration \tcb{$T_*$} has been found and the corresponding controls pulse $c_*(t)$ satisfies the hardware bounds $|c_*(t)|\leq b_{max}$. Note that as $s\to 1$, the scaled control pulses provide better and better initial guesses for solving the optimization problem. As a result, fewer and fewer optimization iterations are required.

In the proposed time-scaling scheme, the gate duration increases whenever the control pulse exceeds the amplitude bound, and decreases whenever the pulse is below the same bound. However, many iterations of the time-scaling scheme would be needed to precisely determine the minimal gate duration \tcb{such that the maximum control amplitude equals the prescribed bound}. To obtain a practically useful stopping criteria, we therefore introduce a \textit{range} of acceptable maximum control amplitudes $[b_{max}-\delta_b, b_{max}]$, for $\delta_b >0$.
The time-scaling iteration then terminates if $c_{max}\in [b_{max}-\delta_b, b_{max}]$; otherwise the gate duration and control coefficients are re-scaled as described above. The resulting time-scaling iteration is summarized in Algorithm~\ref{alg:mintime}. 
\begin{algorithm}
\caption{Minimizing the gate duration}\label{alg:mintime}
\begin{algorithmic}[1]
\Require Specify the range of acceptable maximum control amplitudes: $[b_{max} - \delta_b, b_{max}]$.
\Require Pick an initial gate duration \tcb{$T_0>0$} and select an initial guess for the control vector \tcb{$\vec{\alpha}_0$}.
\For{$k=0,1,2,\dots$}
\State \tcb{Given} $\vec{\alpha}_k$, solve the optimization problem 
\[
   \vec{\alpha}_* = \arg\min_{\vec{\alpha}} J_{\mathrm{infid}}\left(U(T_k;\vec{\alpha})\right) + \gamma J_{\mathrm{energy}}\left(c(\cdot; \vec{\alpha})\right)
\]
   \tcb{such that} $U(T_k;\vec{\alpha})$ solves $\dot U(t;\vec{\alpha}) = -iH(t;\vec{\alpha})U(t;\vec{\alpha})$ with $U(0;\vec{\alpha})=I$.
\If{$\max_{t\in [0,T_k]} \,|c(t, \vec{\alpha}_*)| \in [b_{max} - \delta_b, b_{max}]$}
    \State Success!
    \State \Return $T_k, \vec{\alpha}_*$
\Else 
    %\State Evaluate the scaling factor $s = \max_t|d(t, \alpha_*)|/b_{max}$
    \State Update the gate duration and the initial guess for next control vector: 
        \begin{align*}
            T_{k+1} = sT_k, \quad \text{and} \quad  \vec{\alpha}_{k+1} = \frac{1}{s}\vec{\alpha}_*, \\
            \text{where} \quad s = \frac{\max_{t\in[0,T_k]}|c(t; \vec{\alpha}_*)|}{b_{max}}
        \end{align*}  
\EndIf
\EndFor
\end{algorithmic}
\end{algorithm}

%%%%%%%%%%%%%%%%%%%%%%%%%%%%%%%%%%%%%%%%%%%%%%%%%%%%
\section{Numerical results} \label{sec:results}
In this section, we verify on multiple test cases that the iterative time-scaling scheme yields gate durations that are close to the smallest possible time in which the target gate can be realized, when hardware constraints are in place on  the control pulse amplitude. We here choose a maximum allowable amplitude bound of $b_{max}{/2\pi} = 40$ MHz in the rotating frame, and accept the resulting control pulses (and final times) whenever the maximum control pulse amplitudes are within the range of $[35,40]$ MHz.
The test cases are described in Table \ref{tab:testcases}, consisting of two single-qudit cases with $3$ and $4$ energy levels, one two-qubit case with $2$ energy levels per qubit, and two three-qubit cases with 2 energy levels each. The qubits are coupled to their nearest neighbor(s) in a one-dimensional chain with \tcb{dipole-dipole}  coupling strength $J=5$ MHz. System frequencies for each test case are shown in Table \ref{tab:systemparams}. 
\begin{table}[htb]
 \centering
 \begin{tabular}{@ { } l l l @ { }}
    \toprule
      Name  &   System &  Target gate \\
   \midrule
      QFT$_4$ & one qudit, 4 levels   &  
      $\tcb{ 
      \frac 1 2 \begin{pmatrix} 
        1 & 1 & 1 & 1\\
        1 & $i$ & $-1$ & $-i$\\
        1 & $-1$ & $1$ & $-1$\\
        1 & $-i$ & $-1$ & $i$\end{pmatrix}\in \C^{4\times 4}}$      \vspace{2mm}\\
   % \midrule
      SWAP02 & one qudit, 3 levels & $\begin{pmatrix} 
          0 & 0 & 1 \\  
          0 & 1 & 0 \\  
          1 & 0 & 0 \\  
        \end{pmatrix} \in \R^{3\times 3}$ \vspace{2mm} \\
      CNOT & \begin{tabular}{@{}l@{}}two qubits\\ coupling $1\leftrightarrow 2$ at 5MHz \end{tabular} & $\begin{pmatrix} 
          1 & 0 & 0 & 0 \\  
          0 & 1 & 0 & 0 \\  
          0 & 0 & 0 & 1 \\  
          0 & 0 & 1 & 0 \  
        \end{pmatrix} \in \R^{4\times 4}$ \vspace{2mm} \\
      CCNOT &  \begin{tabular}{@{}l@{}}three-qubit chain\\ coupling $1\leftrightarrow 2$ at 5MHz \\ coupling $2\leftrightarrow 3$ at 5MHz\end{tabular}  &  $\begin{pmatrix} 
          1  \\  
            & \ddots   \\  
            &        & 1  \\  
            &        &   & 0 & 1 \\  
            &        &   & 1 & 0 
        \end{pmatrix} \in \R^{8\times 8}$ \vspace{2mm}\\
      SWAP chain& \begin{tabular}{@{}l@{}}three-qubit chain\\ coupling $1\leftrightarrow 2$ at 5MHz \\ coupling $2\leftrightarrow 3$ at 5MHz \end{tabular} & $\begin{pmatrix} 
          1  \\  
            & 0 &   &   & 1 \\  
            &   & 1 &   \\  
            &   &   & 0 &  & & 1 \\  
            & 1  &  &   & 0 \\  
            &    &  &   &   & 1 \\  
            &    &  & 1 &   &   & 0 \\  
            &    &  &   &   &   &  & 1 \\  
        \end{pmatrix}  \in \R^{8\times 8} $ \\
    \bottomrule
  \end{tabular} 
  \vspace{2mm}
  \caption{Test case specifications and target gates.}
  \label{tab:testcases}
\end{table}
\begin{table}[htb]
\begin{adjustbox}{max width=1.1\textwidth,center}
  \centering
 \begin{tabular}{@ { } l  l | l l l l l @ { }}
    \toprule
      Name  & $dim(H)$ & $\omega_q/2\pi$ [GHz] & $\xi_q/2\pi$ [GHz] & $J_{pq}/2\pi$ [GHz] & $\omega^{rot}/2\pi$ [GHz]  & $\Delta_{B}$ [ns] \\
    \midrule
      QFT$_4$ & $d=4$ & 4.914 & 0.33 & - & 4.584 & 0.3 \\
      SWAP02 & $d=3$ & 5.12 & 0.34 & - & 4.78 & 0.3\\
      CNOT & $d=2\times 2$ & 5.12, 5.06 & 0.34, 0.34 & 0.005 & 5.09 & 1.65 \\
      CCNOT & $d=2\times 2\times 2$ & 5.18, 5.12, 5.06 & 0.34, 0.34, 0.34 & 0.005, 0.0, 0.005 & 5.12 & 1.65 \\
      SWAP chain& $d=2\times 2\times 2$ & 5.18, 5.12, 5.06 & 0.34, 0.34, 0.34 & 0.005, 0.0, 0.005 & 5.12 & 1.65\\
    \bottomrule
  \end{tabular} 
 \end{adjustbox}
 \vspace{2mm}
  \caption{System parameters.}
  \label{tab:systemparams}
\end{table}

\begin{figure}[htb]
    \centering
    \includegraphics[width=0.5\textwidth]{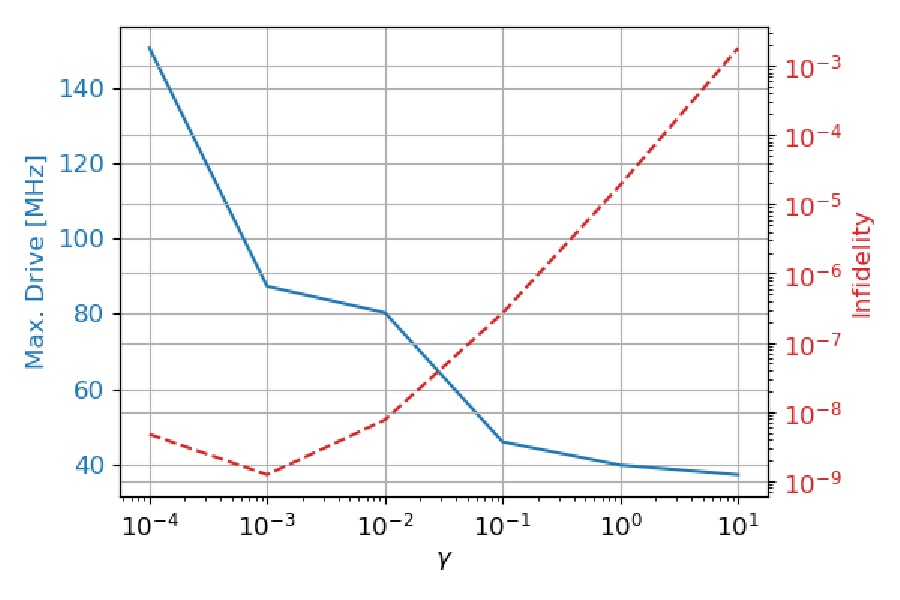}
    \caption{\tcb{Maximum control drive amplitude and infidelity when optimizing for the $QFT_4$ gate for various values of the tunable parameter $\gamma$ in the objective function \eqref{eq:optimproblem}. Here, the gate duration is fixed to $T=20$ ns.}}
    \label{fig:dmax_varying_gamma}
\end{figure}

We solve the inner optimization problem in Algorithm \ref{alg:mintime} using L-BFGS as implemented in the quantum control software Quandary \cite{gunther2021quandary}. 
In order to ensure that the inner optimization cycle generates control pulses with minimal energy norm, we choose the penalty coefficient to be $\gamma = 1$.
\tcb{While $\gamma$ can be considered a tunable parameter in the algorithm, we motivate this value of $\gamma$ through a numerical experiment. Here we study the effect of varying $\gamma$ on the maximum control drive amplitude and infidelity that results from minimizing the objective \eqref{eq:optimproblem}. Because the optimization minimizes the sum of two terms, it is to be expected that increasing $\gamma$ would put more emphasis on minimizing the control energy and less emphasis on the infidelity. Indeed, the results shown in Fig.~\ref{fig:dmax_varying_gamma} verify the expected behavior. In the numerical tests presented below we aim the final infidelity to be less than $10^{-4}$. In the case tested here, $\gamma=1$ meets the criteria for the infidelity, while also significantly reducing the maximum control drive amplitude.}

In order to ensure that an optimal point with minimal energy norm has been reached, we note that it is important to base the optimization stopping criterion on the norm of the gradient, rather than on a small gate infidelity. \tcb{In the following, we set the gradient threshold to be $10^{-5}$.} 
To stabilize the inner optimization convergence, we further add a Tikhonov regularization term \tcb{$\gamma_1\|\vec\alpha\|^2$}, with parameter $\gamma_{1} = 10^{-2}$.
\tcb{The purpose of the Tikhonov term is to regularize the optimization problem by adding a convex term that shifts the eigenvalues of the Hessian of the objective function by a small constant factor $\gamma_1$. This factor should be chosen small enough to not alter the optimization outcome \cite{nocedal2006numerical}. We remark that, when the controls are parameterized by B-spline basis functions, the control energy term \eqref{eq:energynorm} can be interpreted as a weighted Tikhonov regularization term, see Appendix~\ref{app:energypenalty_regul}.} 
\tcb{In addition, to encourage a state evolution where the populations vary slowly in time, we also penalize the second time-derivative of the state populations. To this end, we add a penalty term of the form}
\begin{align}
\tcb{\frac{\gamma_2}{2T}\int_0^T \sum_{i,j} \left( \frac{\partial^2}{\partial^2 t} |U_{i,j}(t)|^2 \right)^2  \, \mathrm{d}t,}
\end{align}
\tcb{where $U_{i,j}$ denotes element $(i,j)$ in the matrix $U$.}
\tcb{Again, $\gamma_2\geq 0$ should be chosen small enough to not dominate the  infidelity and the control energy terms during the optimization. In the following, we use $\gamma_{2} = 10^{-2}$.
While the qualitative results of the minimal-gate-duration scheme are not expected to be sensitive to the precise value of $\gamma_2$, a thorough analysis of this term remains to be done.} 

During the initialization phase of Algorithm \ref{alg:mintime} ($k=0$), we assign the elements of the initial control vector from a uniform probability distribution $\vec{\alpha}_0\sim {\cal U}(-0.9 b_{max},0.9 b_{max})$.

\begin{figure}[p]
% \begin{figure*}[p]
    % \vspace*{-1cm}
    \centering
    \begin{subfigure}[b]{\textwidth}
       \centering
       \includegraphics[width=0.40\textwidth]{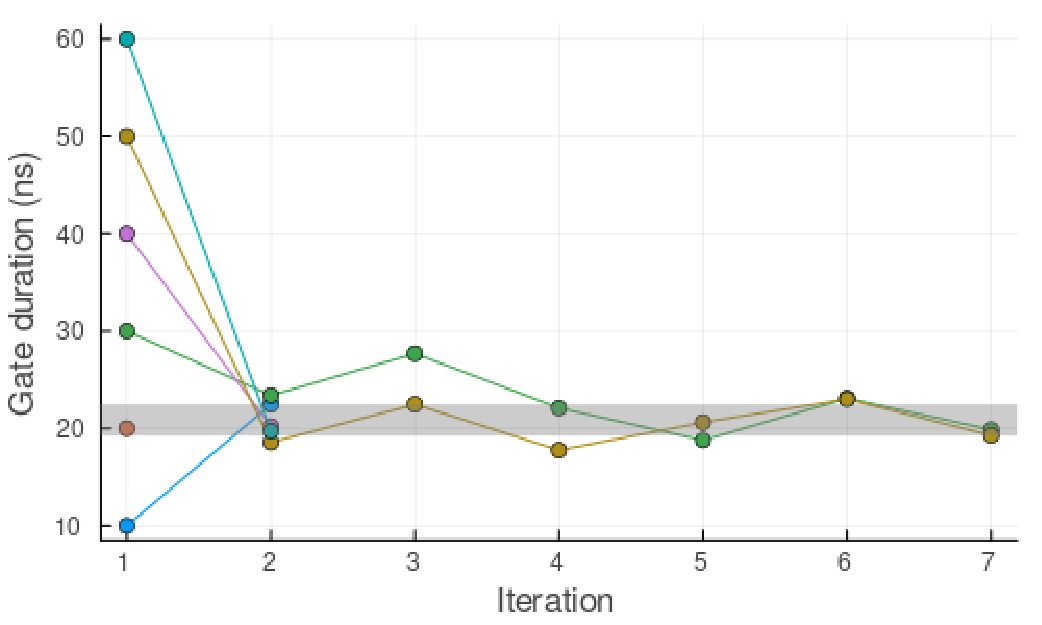}
       \hspace{1cm}
       \includegraphics[width=0.40\textwidth]{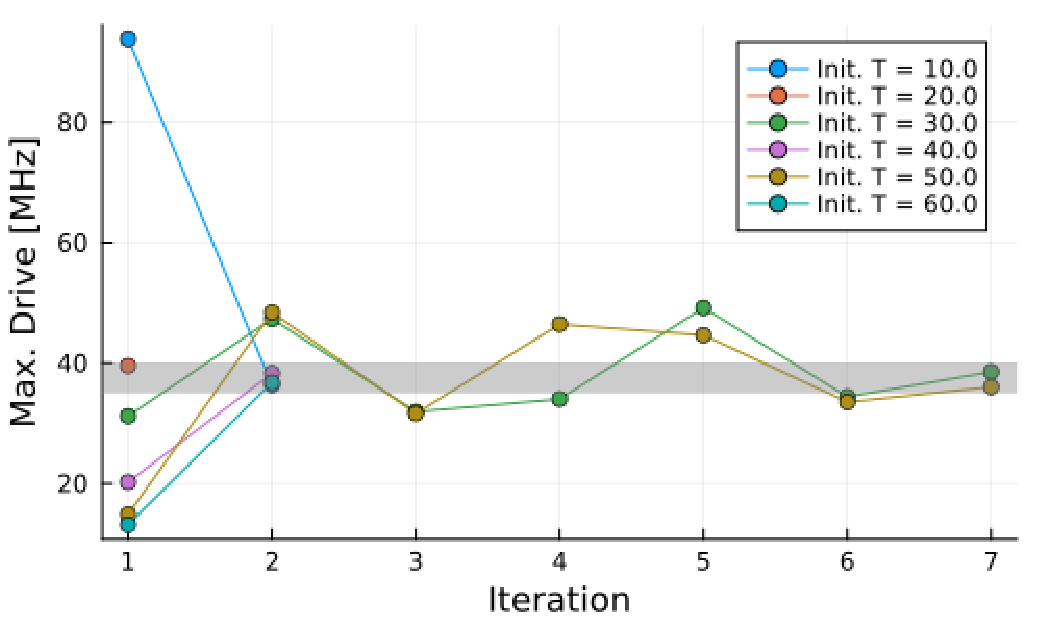}
        \caption{QFT$_4$ gate on a single qubit.}
        \label{fig:QFT}
     \end{subfigure}
     \begin{subfigure}[b]{\textwidth}
       \centering
        \includegraphics[width=0.40\textwidth]{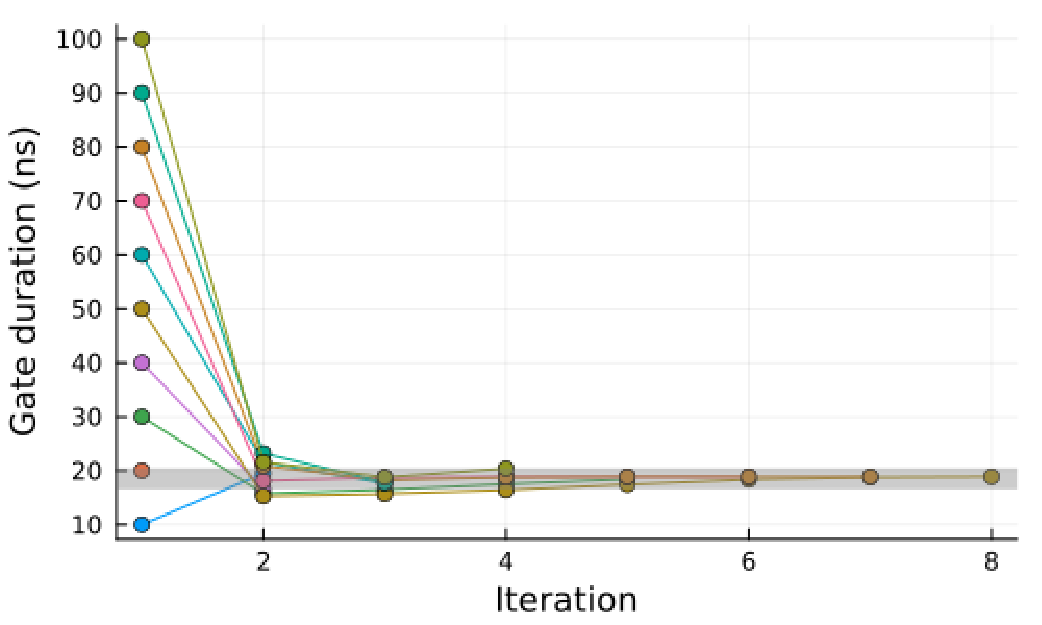}
        \hspace{1cm}
        \includegraphics[width=0.40\textwidth]{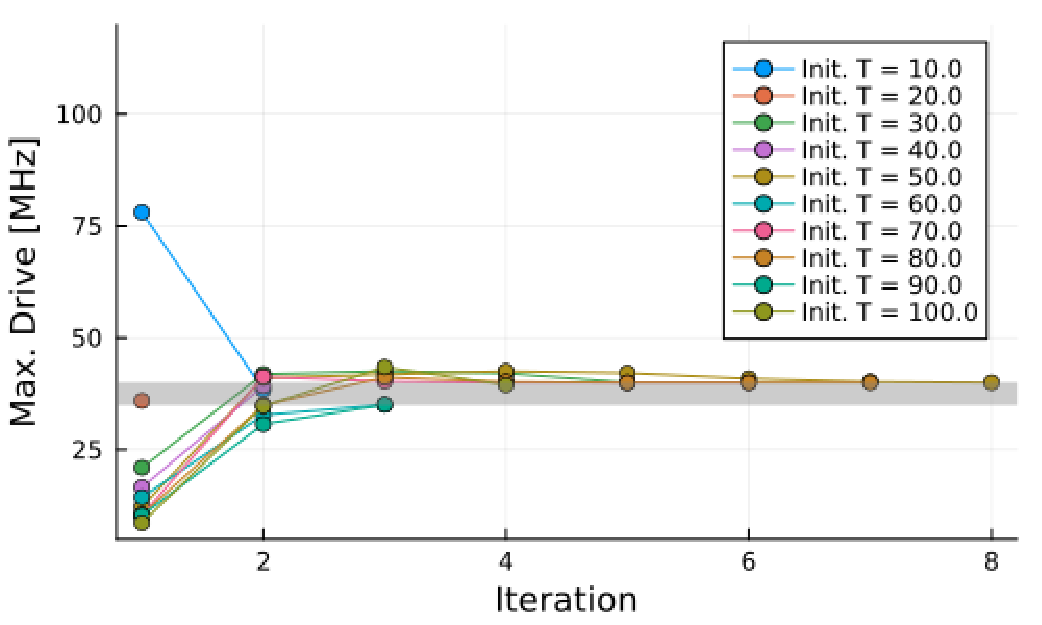}
        \caption{SWAP02 gate on a single qubit}
        \label{fig:SWAP02}
     \end{subfigure}
     \begin{subfigure}[b]{\textwidth}
       \centering
       \includegraphics[width=0.40\textwidth]{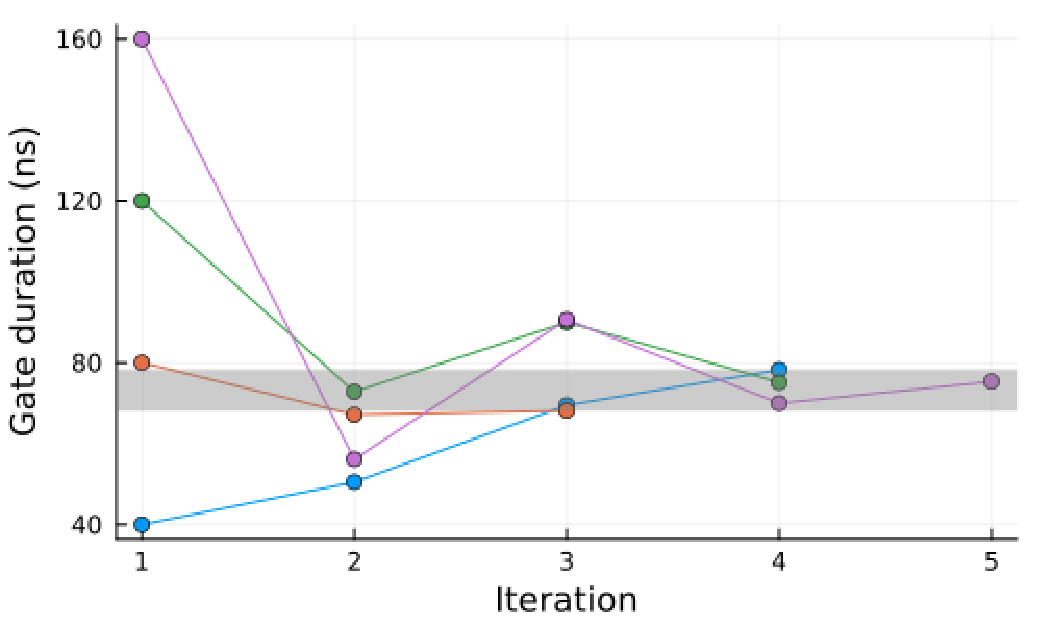}
        \hspace{1cm}
       \includegraphics[width=0.40\textwidth]{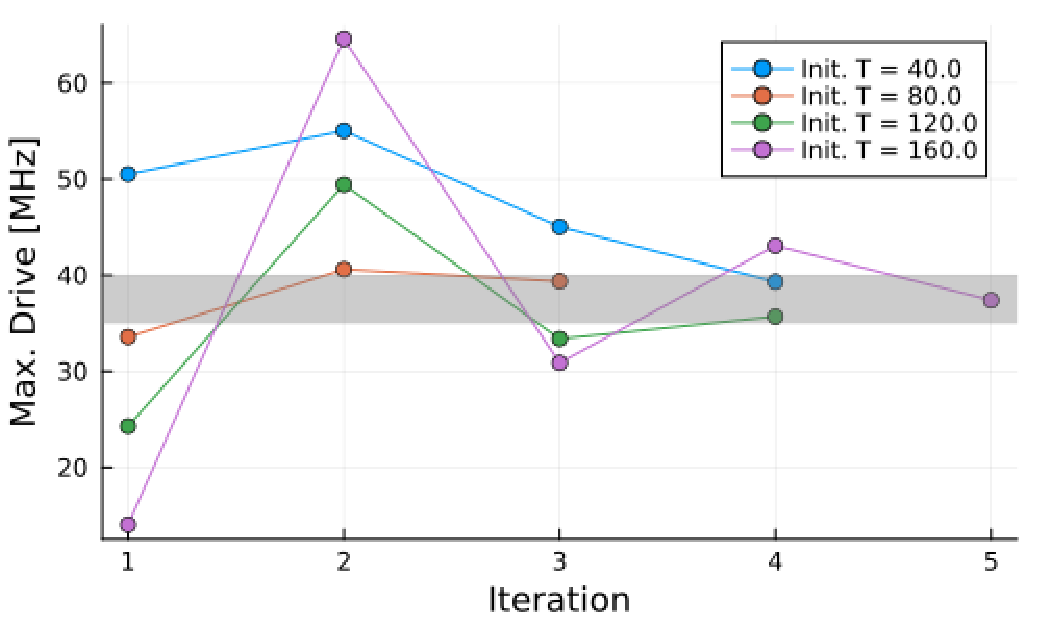}
       \caption{CNOT gate, coupling $J_{12}=5$MHz}
        \label{fig:CNOT}
     \end{subfigure}
     \begin{subfigure}[b]{\textwidth}
       \centering
       \includegraphics[width=0.40\textwidth]{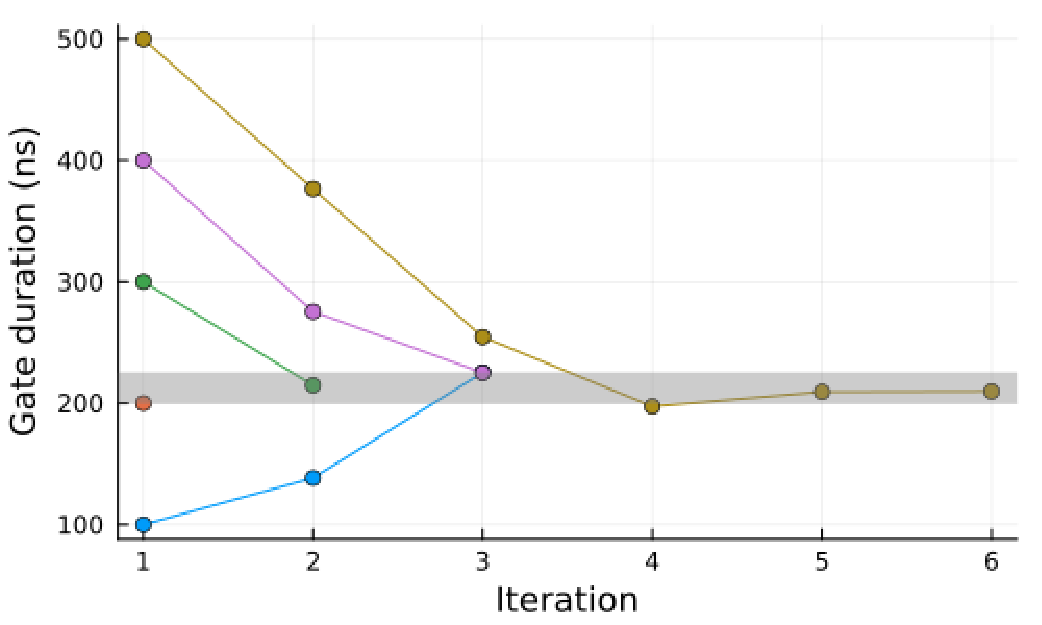}
       \hspace{1cm}
       \includegraphics[width=0.40\textwidth]{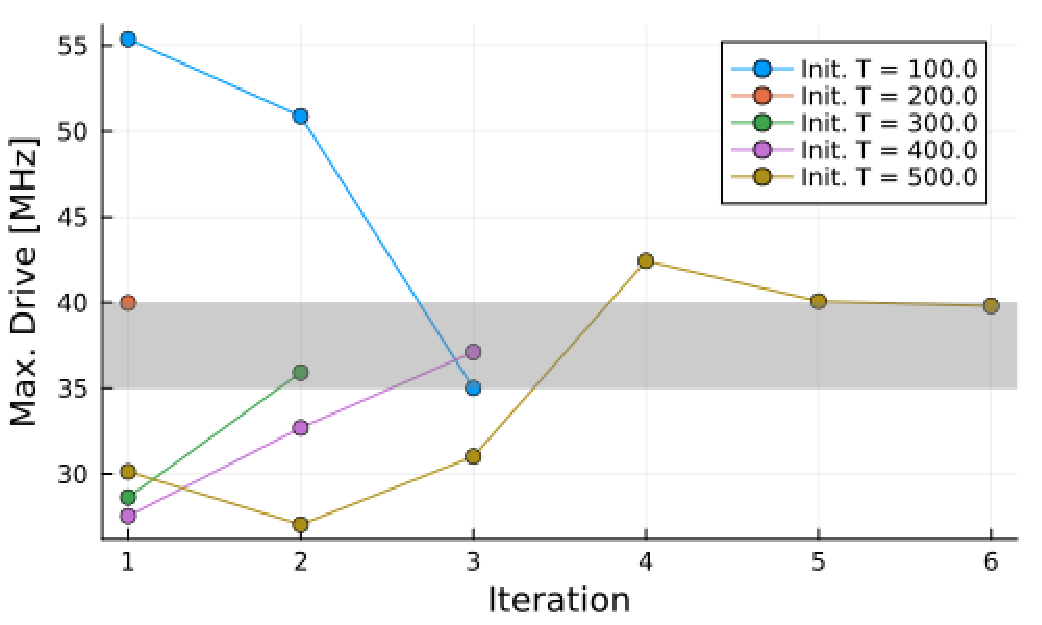}
       \caption{CCNOT (Toffoli) gate on a three-qubit chain}
       \label{fig:C2NOT}
     \end{subfigure}
     \begin{subfigure}[b]{\textwidth}
       \centering
       \includegraphics[width=0.40\textwidth]{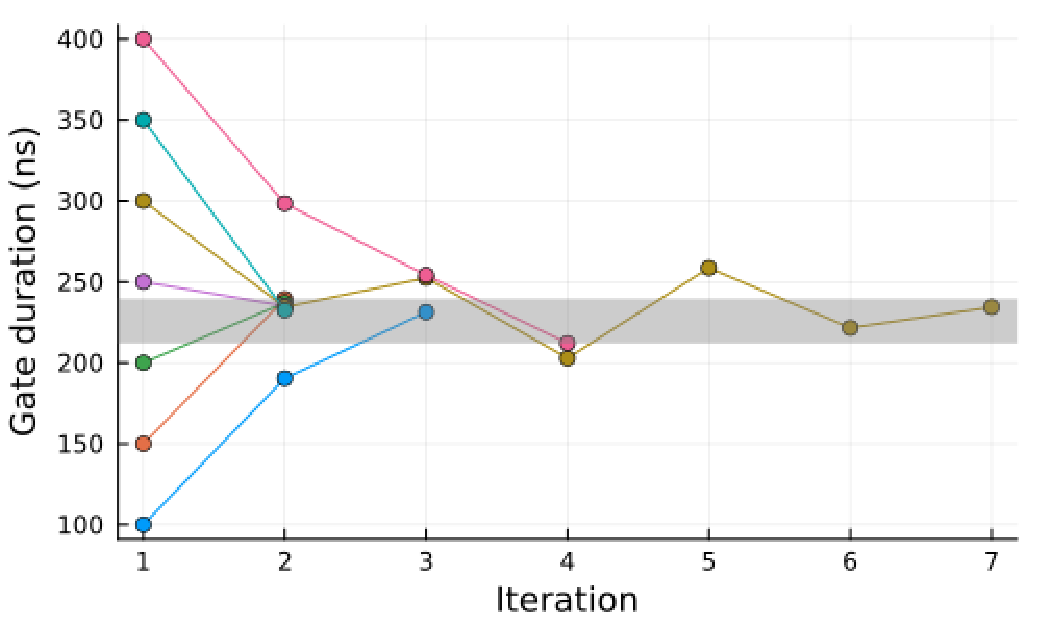}
       \hspace{1cm}
       \includegraphics[width=0.40\textwidth]{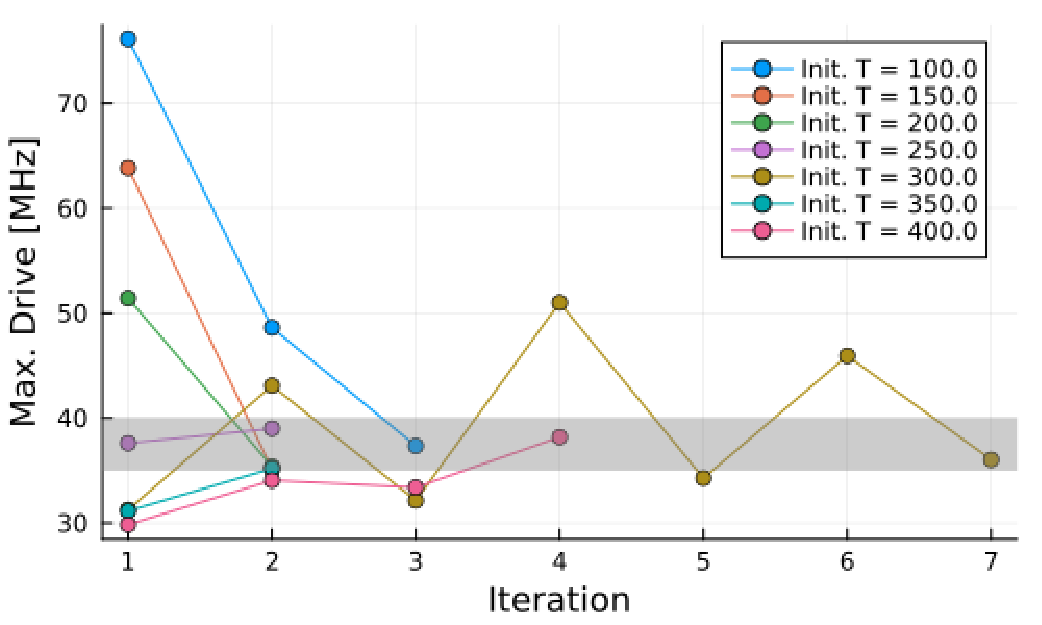}
      \caption{SWAP qubit 1 and 3 on a three-qubit chain}
     \label{fig:SWAP_threequbit_mintime}
     \end{subfigure}
     \caption{Optimization history for the time-scaling algorithm (Alg.~\ref{alg:mintime}): Gate durations (left column) and maximum amplitudes (right column). The gray areas denote the resulting band of final gates times, and the prescribed range of acceptable amplitudes, respectively.}
     \label{fig:mintime_results}
% \end{figure*}
\end{figure}

For each test case, we deploy Algorithm \ref{alg:mintime} for a wide range of initial gate durations, \tcb{$T_0$}. Figure \ref{fig:mintime_results} shows the convergence history of the intermediate gate durations $T_k$ (left column), as well as the corresponding maximum \tcb{drive amplitude of the optimized pulses (right column) at each iteration}.
Across all test cases, we observe that an optimal gate duration can be found within very few iterations (often only two, but never more than 8 iterations), meaning that only a small number of optimal control cycles are required to determine the minimal gate duration, for a given bound on the control amplitude. This provides a clear advantage over the standard (brute-force) approach of sweeping over a wide range of gate durations, which can easily require tens to hundreds of optimal control cycles to be completed. 
The resulting final gate times at the last cycle of the time-scaling scheme are indicated by gray bands in Figure~\ref{fig:mintime_results}, and are also provided in Table~\ref{tab:mintime_vs_fromscratch}. The width of these bands is determined from the user-prescribed range of allowable maximum amplitudes $[b_{max}-\delta_b, b_{max}]$, \tcb{which serves as the stopping criterion for the minimal time iteration.} 
\tcb{Note that we here consider $b_{\mathrm{max}} = 40$MHz and $\delta_b=5$MHz, allowing for a $12.5\%$ deviation from the maximum bound. The variations of final minimal times as presented in Table~\ref{tab:mintime_vs_fromscratch} are of the same order. Importantly, we observe that the iterations converge to this band for a wide range of initial gate durations, $T_0$.}
%

% \AddThispageHook{\thispagestyle{empty}} % Remove page number and header 
% }

\begin{figure}[htb]
    \centering
    \begin{subfigure}[b]{0.40\textwidth}
        \centering
        \includegraphics[width=\textwidth]{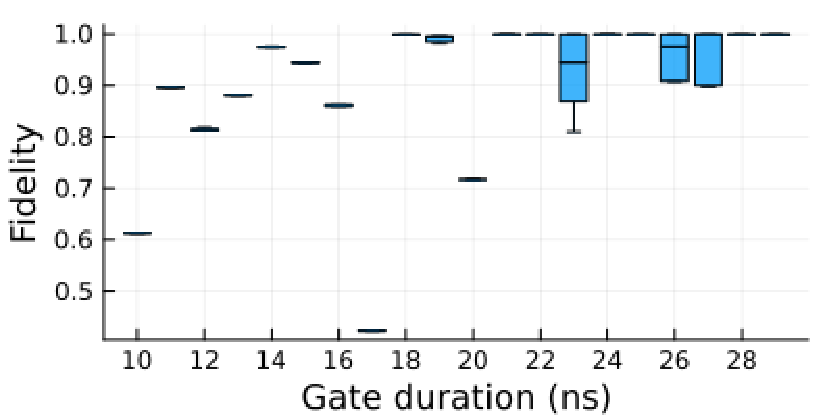}
        \caption{QFT$_4$ gate on single qubit.}
     \end{subfigure}
     \hspace{1cm} 
     \begin{subfigure}[b]{0.40\textwidth}
        \includegraphics[width=\textwidth]{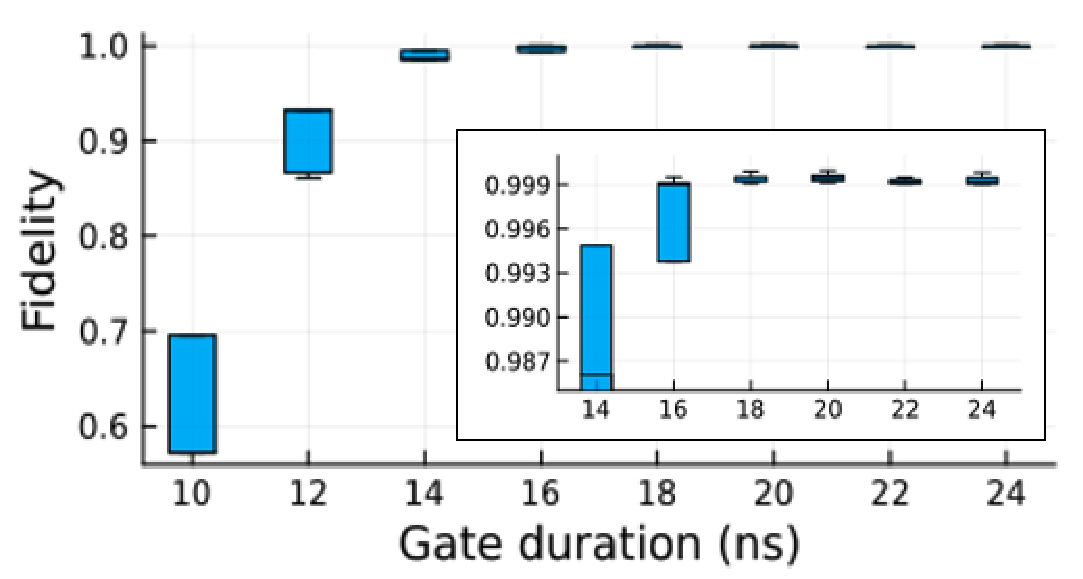}
        \caption{SWAP02 gate on single qubit}
     \end{subfigure}
     \hspace{1cm} 
     \begin{subfigure}[b]{0.40\textwidth}
       \includegraphics[width=\textwidth]{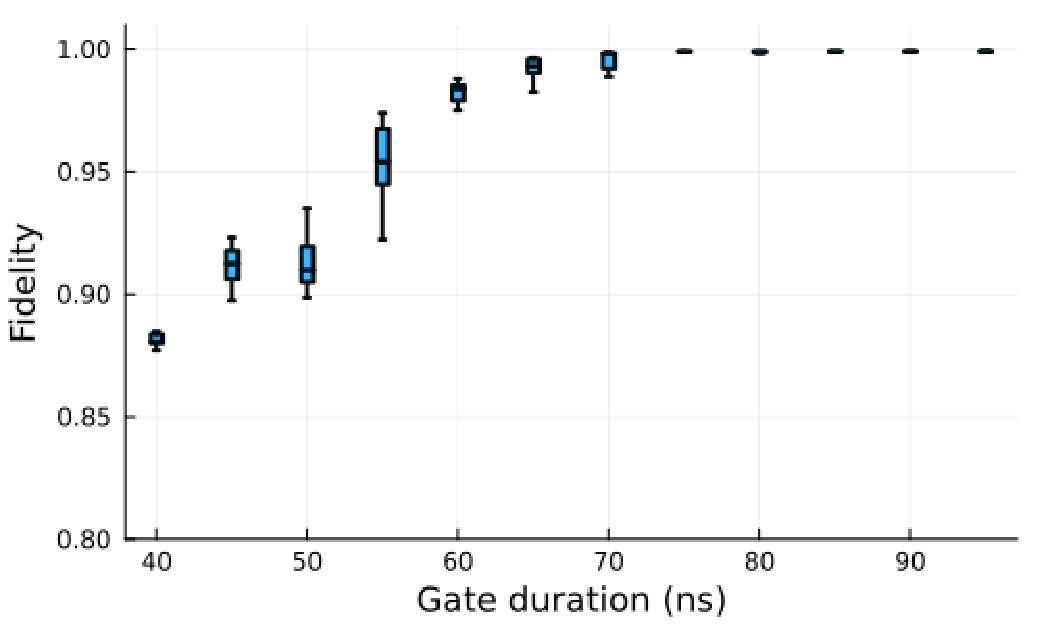}
       \caption{CNOT gate, coupling $J_{12}=5$MHz}
     \end{subfigure}
     \hspace{1cm} 
     \begin{subfigure}[b]{0.40\textwidth}
       \includegraphics[width=\textwidth]{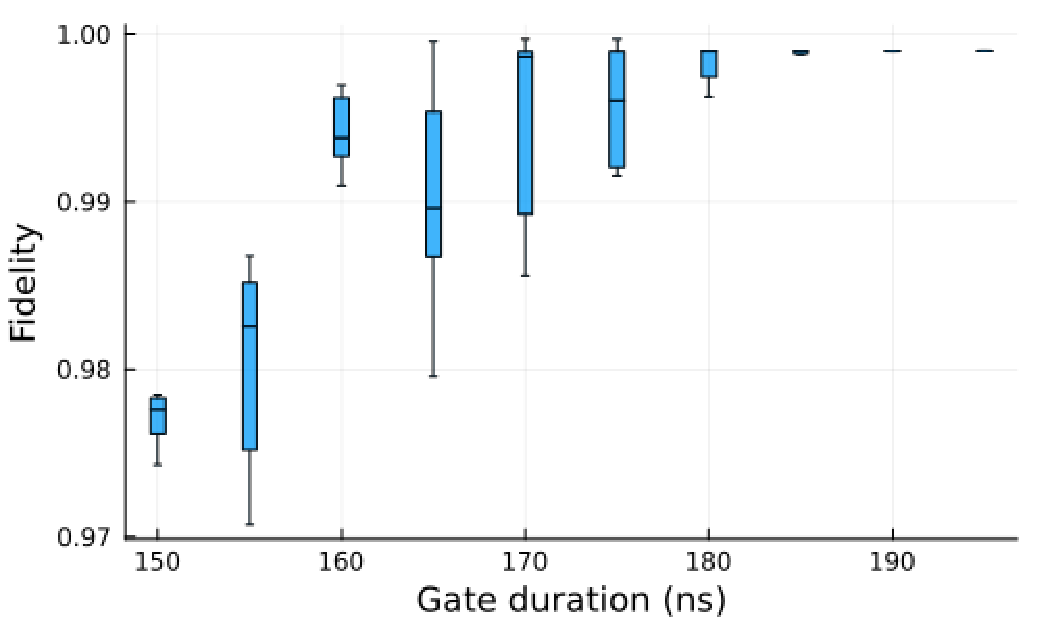}
       \caption{Toffoli gate on a three-qubits chain.}
     \end{subfigure}
     \hspace{1cm} 
     \begin{subfigure}[b]{0.40\textwidth}
       \includegraphics[width=\textwidth]{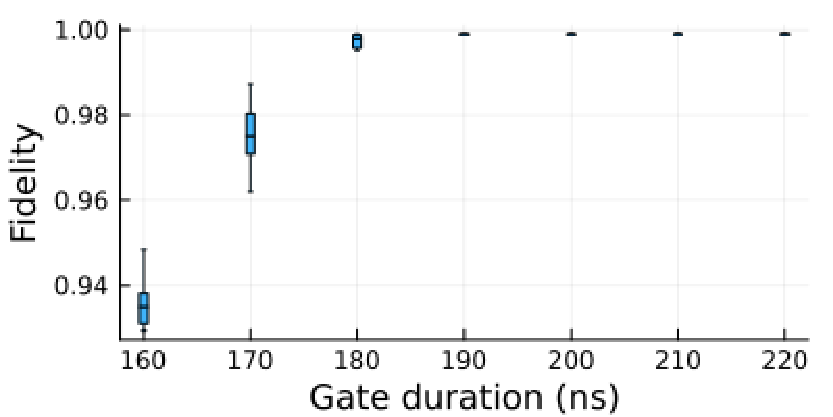}
        \caption{SWAP qubit 1 and 3 on a three-qubit chain}
     \end{subfigure}
    \caption{Estimating the quantum speed limit with the brute-force method: Optimized fidelities sweeping over various gate durations, while enforcing hardware constraints directly during the optimization process. For each gate duration, statistics are gathered by performing 10 optimization cycles starting from randomly sampled initial control vectors.}
    \label{fig:optim_from_scratch}
\end{figure}

To evaluate how well the iterative time-scaling scheme recovers the actual minimal gate duration for a given amplitude bound, 
we compare the resulting gate durations with a brute-force approach that performs a sweep of optimizations for a wide range of (fixed) gate durations, while enforcing box constraints on the control vector using a projected L-BFGS method. Figure \ref{fig:optim_from_scratch} shows statistics of optimized gate fidelities, each gathered over 10 different optimizations starting from random initial control vectors. Here, variations in the resulting gate fidelities can be attributed to deterioration of the optimization progress, related to the imposed box-constraints. Nevertheless, we compare the resulting minimal gate durations as observed in Figure \ref{fig:optim_from_scratch} with those produced by Algorithm \ref{alg:mintime} as in Figure \ref{fig:mintime_results}.  Table \ref{tab:mintime_vs_fromscratch} summarizes the resulting gate durations, demonstrating that the time-scaling scheme converges to final gate durations that are remarkably close to those obtained with the brute force method. The range of resulting gate durations as produced by Algorithm \ref{alg:mintime} is a direct consequence of the stopping criteria based on a prescribed range of allowed amplitudes $[b_{max} - \delta_b, b_{max}]$. The range in gate duration hence depends on the user-controlled parameter $\delta_b$. In the last column, we compare the shortest gate durations with the ones obtained from the brute-force optimization approach. We observe a relative difference below $7\%$ for all test cases.  

\begin{table}[htb]
 \centering
 \begin{tabular}{@ { } l | l l | c @ { }}
    \toprule
      Case  & Algorithm~\ref{alg:mintime} & Brute force & Smallest rel.~diff.\\
   \midrule
      QFT$_4$ & 19ns to 23ns & 18ns to 21ns & 5\%\\
      SWAP02 & 18ns to 23ns & 18ns & 0\%\\
      CNOT & 68ns to 78ns & 70ns & -2\% \\
      CCNOT & 200ns to 225ns & 190ns & 5\% \\
      SWAP chain & 205ns to 239ns & 190ns & 7\% \\
    \bottomrule
  \end{tabular}
  \vspace{2mm}
  \caption{Minimal gate durations with gate fidelity $\geq 99.9\%$, as achieved from the time-scaling scheme in Algorithm \ref{alg:mintime} (Figure \ref{fig:mintime_results}), and brute-force optimization with box constraints on the control vector (Figure \ref{fig:optim_from_scratch}).}
 \label{tab:mintime_vs_fromscratch}
\end{table}

\section{Conclusion}\label{sec:conclusion}
This paper presents a practical scheme for approximating the minimal duration for realizing a unitary gate while satisfying given amplitude bounds on the control pulses. The scheme performs a sequence of unconstrained optimal control cycles, each minimizing the gate fidelity alongside an additional penalty term for the control pulse amplitudes. After each cycle, the gate duration is scaled to a new value for which the scaled controls satisfy the amplitude bounds. We provide numerical evidence of convergence to a final gate duration that is close to the shortest achievable duration obtained from sweeping over a large number of gate durations. Updating the gate duration based on the previous control amplitudes converges to the minimal gate duration in a few optimization cycles, drastically reducing the computational costs for practical usage of time-optimal quantum control, which often requires tens to hundreds of optimization cycles to sweep over a range of gate durations in a trial and error fashion.
We demonstrate that the proposed scheme converges for a wide range of initial gate durations, indicating that almost no prior knowledge of the minimal gate duration is required.
The proposed technique is straightforward to be implemented on top of an existing quantum optimal control code, where we note that it is beneficial to parameterize the control pulses by basis functions such that the pulse amplitudes can easily be adjusted. Further, the proposed method is agnostic to the underlying system and control Hamiltonian models, as well as the target unitary gate operation, making the time-scaling iteration an easy to implement and practically useful scheme for reducing the durations of quantum gate operations.

As illustrated in Figure \ref{fig:H4_nobounds_dmax}, the control pulse amplitude can be a non-monotonic function of the gate duration. We conjecture that this phenomena is the underlying cause of the non-monotonic convergence (over- and under-shooting the gate duration) of our algorithm, and we conjecture that convergence may be further improved by damping the updates of the duration. This will be subject of further investigations.

%%%%%%%%%%%%%%%%%%%%%%%%%%%%%%%%%%%%%%%%%%%%%%%%%%%%
\section*{Acknowledgments}
The authors would like to acknowledge Dr.~Jonathan Dubois at Lawrence Livermore National Laboratory, CA, who provided expertise and partial funding resources during the preparation of this paper.
% \section*{Auspices}

This work was performed under the auspices of the U.S. Department of Energy by Lawrence Livermore National Laboratory under Contract DE-AC52-07NA27344. LLNL-JRNL-851967.

\appendix

%%%%%%%%%%%%%%%%%%%%%%%%%%%%%%%%%%%%%%%%%%%%%%%%%%%%
\section{The infidelity is invariant to the control pulse integral in a qubit model problem}

\label{app:displacement}

To gain further insight into the essential properties of a control pulse, we
consider the model problem of a driven two-level system {in the rotating frame}, 
\begin{equation}
    H(t) = c(t) a^\dagger + c^*(t) a,\quad c(t): \mathbb{R} \mapsto \mathbb{C}.
\end{equation}
We consider the displacement transformation~\cite{GerryKnight-05},
\begin{equation}
    U(t) = D(\beta(t)) \widetilde{U}(t),\quad \beta(t):\,\mathbb{R}\mapsto \mathbb{C},
\end{equation}
where the unitary transformation operator is
\begin{equation}
    D(\beta) = \exp(\beta a^\dagger - \beta^* a),\quad D(0) = I.
\end{equation}
Thus, by taking $\beta(0)=0$, we have $U(0) = \widetilde{U}(0)$ and the transformed unitary solution operator satisfies
\begin{align}
    \dot{\widetilde{U}}(t) &= -i \widetilde{H}(t)\widetilde{U}(t),\quad \widetilde{U}(0) = I,\\
    \widetilde{H}(t) &= \left( c(t) - i \frac{d\beta(t)}{dt}\right)a^\dagger + \left( c^*(t) + i \frac{d\beta^*(t)}{dt}\right)a,
\end{align}
where we have omitted terms of the form $\delta(t)I$ in $\widetilde{H}(t)$ because they only change the global phase in $\widetilde{U}$. We conclude that $\widetilde{U}(t) = I$, for all $t\geq 0$, if $\beta(t)$ satisfies the ordinary differential equation
\begin{equation}
    \frac{d\beta(t)}{dt} = - ic(t),\ %
    \beta(0)=0,\ \Leftrightarrow\ % 
    \beta(t) = - i \int_0^t c(\tau) \, d\tau.
\end{equation}
In this case, we have $U(T) = D(\beta(T))\widetilde{U}(T) = D(\beta(T))$, {and} the infidelity at time $t=T$ satisfies 
\begin{equation}
    J_{\mathrm{infid}} = 1 - \left| \frac{1}{N} \mbox{Tr}\left( V_{\textrm{target}}^\dagger D(\beta(T))\right)
    \right|^2.
\end{equation}
{For a given duration $T$, we note that the infidelity only depends on the integral $\beta(T) = -i\int_0^T c(\tau) \, d\tau$. As a result, any control function $\widetilde{c}(t)$ with the same integral over time yields the same infidelity.}

If we scale time, $t' = s t$, and define $\beta'(t') = \beta(t)$, the infidelity with respect to $V_{\textrm{target}}$ becomes invariant to the scaling as long as $\beta'(T') = \beta(T)$. The scaling implies
\begin{equation}
    \beta'(T') = -i\int_0^{T'} c'(\tau') \, d\tau' 
    = -i\int_0^T c'(s \tau) s\, d\tau.
\end{equation}
Hence, $\beta'(s T) = \beta(T)$ if the scaled control pulse satisfies
\begin{equation}
    c'(s t) = \frac{c(t)}{s}.
\end{equation}
For example, if $s>1$ the duration of the control pulse is increased and its amplitude is decreased, such that the time integral of the control pulse remains unchanged. 

%%%%%%%%%%%%%%%%%%%%%%%%%%%%%%%%%%%%%
\section{\tcb{Control parameterization via B-spline basis functions}}\label{app:bspline}

\tcb{The control function $c(t)$ is parameterized in terms of a weighted sum of B-spline basis functions, according to \eqref{eq:bsplinebasis}. It is convenient to split the complex-valued function into its real and imaginary parts, $c(t) = p(t) + i q(t)$, where
\begin{align}
    p(t) = \sum_{s=1}^{N_s} \alpha^{real}_s B_s(t), \quad q(t) = \sum_{s=1}^{N_s} \alpha^{imag}_s B_s(t).
\end{align}
Here, $N_s\geq 1$ equals the number of terms in $p(t)$ and $q(t)$, specified by the real-valued design variables $\{\alpha^{real}_s, \alpha^{imag}_s \}_{s=1}^{N_s}$. In this work we let $B_s(t)$ be quadratic B-spline basis functions centered on a uniform grid in time: $t_s = (s+0.5)$\tcb{$\Delta_B$}, with knot spacing \tcb{$\Delta_B$}$= T/(N_s + 2)$.
Each basis function follows by shifting a normalized wavelet function in time, $B_s(t)=\widetilde{B}(\tau_s(t))$, where $\tau_s(t) = (t-t_s)/3$\tcb{$\Delta_B$} and
\begin{align}\label{eq:splinebasis}
  \widetilde{B}(\tau) &= \begin{cases}
    \frac{9}{8} + \frac{9}{2} \tau + \frac{9}{2} \tau^2, \quad & -\frac{1}{2} \leq \tau < -\frac{1}{6}, \\
    \frac{3}{4} - 9 \tau^2, \quad  & -\frac{1}{6} \leq \tau < \frac{1}{6}, \\
    \frac{9}{8} - \frac{9}{2} \tau + \frac{9}{2} \tau^2, \quad & \hphantom{-} \frac{1}{6} \leq \tau < \frac{1}{2}, \\
    0, \quad & \mbox{otherwise}.
  \end{cases}
\end{align}
Note that $B_s(t)$ has compact support in $t\in[t_s - 1.5\Delta_B, t_s + 1.5\Delta_B]$. This implies that, at any time $t\in[0,T]$, at most three terms are non-zero in $p(t)$ and $q(t)$. Since the first and last basis functions are centered at $t_1=3\Delta_B/2$ and $t_{N_s} = T - 3\Delta_B/2$, respectively, the control function always starts and ends with zero amplitude.} 

\tcb{We note that the quadratic B-spline basis functions used here are continuously differentiable.
Control pulses that are only piece-wise continuous, such as those generated by a bang-bang ansatz, can only be \textit{approximated} by a quadratic (second order) B-spline function. If piece-wise constant control pulses are desirable, they could instead be parameterized by zeroth order B-spline functions, where the basis consists of box-car functions.
For a general introduction to B-spline basis functions, compare  ~\cite{PiegTill96}.}
\tcb{We remark that the minimal time algorithm as presented in Section \ref{sec:mintime} is agnostic to the underlying parameterization, as long as the generated pulses can be stretched or compressed to a different pulse length.}

\section{\tcb{Energy penalty term as a regularizer}}\label{app:energypenalty_regul}
\tcb{The integral in the energy penalty term, defined by \eqref{eq:energynorm}, can be written as
\begin{align}
    {\cal J}_{energy} = \frac{1}{T}\int_0^T \left( p(t)^2 + q(t)^2 \right)\, dt.
\end{align}
}
\tcb{We here analyze the regularizing effect this term has on the optimization problem, when the control pulses are parameterized by quadratic B-splines, as defined in Appendix \ref{app:bspline}.
Because each B-spline basis function is a piece-wise quadratic function of time it is straightforward to analytically evaluate the integral in the above formula. After some algebra,
\begin{align}\label{eq:energynorm_app}
    {\cal J}_{energy} = \frac{3\delta}{T} \sum_{j=1}^{N_s} \sum_{k=1}^{N_s} \left(\alpha^r_j W_{j,k} \alpha^r_k + \alpha^i_j W_{j,k} \alpha^i_k\right),
\end{align}
where $W$ is a real, symmetric, penta-diagonal $N_s\times N_s$ matrix,} 
\tcb{
\begin{align}
    W = \frac{11}{60} \begin{bmatrix}
        a & b & c & & & & &\\
        \ddots & \ddots & \ddots & \ddots & & & & &\\
        \ddots & \ddots & \ddots & \ddots & \ddots & & & &\\
        & c & b & a & b & c &  \\
        &  & \ddots & \ddots & \ddots & \ddots & \ddots  \\
        & & & \ddots & \ddots & \ddots & \ddots   \\
        & & & & c & b & a 
    \end{bmatrix},
\end{align}
and the non-zero elements are $a = 1$, $b=\frac{26}{66}$, and $c = \frac{1}{66}$.
The matrix is strictly diagonally dominant for all interior rows, because $a - 2(b + c) = \frac{10}{66} > 0$; the diagonal dominance is even greater for the first and last two rows. Thus, the matrix is positive definite, implying that the energy penalty term also serves to regularize the optimization problem, because \eqref{eq:energynorm_app} represents a weighted norm of the control parameter vector.
}

\bibliographystyle{plain}
\bibliography{mybib.bib}

\end{document}